\documentclass[showpacs,preprintnumbers,amsmath,amssymb]{revtex4}

\usepackage{graphicx}
\usepackage{dcolumn}
\usepackage{bm}
\usepackage{color}
 
\def\mbi#1{\mbox{\bfseries\itshape #1}} 

\begin{document}

\preprint{APS/123-QED}

\title{Effects of a Primordial Magnetic Field on Low and High Multipoles of  the CMB }

\author{Dai G. Yamazaki$^{1}$}
 \homepage{http://th.nao.ac.jp/~yamazaki/}
 \email{yamazaki@th.nao.ac.jp}
\author{Kiyotomo Ichiki$^{2}$}%
\author{Toshitaka Kajino$^{1,3}$}%
\author{Grant J. Mathews$^{4}$}%
\affiliation{%
$^{1}$National Astronomical Observatory, Japan
Mitaka, Tokyo 181-8588, Japan
}%
\affiliation{%
$^{2}$Research Center for the Early Universe,
School of Science, the University of Tokyo,
7-3-1 Hongo, Bunkyo-ku 113-0033, Japan
}%
\affiliation{%
$^{3}$Department of Astronomy, Graduate School of Science, University of Tokyo
7-3-1 Hongo, Bunkyo-ku, Tokyo 113-0033, Japan
}%
\affiliation{%
$^{4}$Center for Astrophysics,
Department of Physics, University of Notre Dame, Notre Dame, IN 46556, U.S.A.
}%

\date{\today}

\begin{abstract}
 The existence of a primordial magnetic field (PMF) would affect both the temperature and polarization
 anisotropies of the cosmic microwave background (CMB).  It also provides a plausible  explanation for the possible disparity  between
 observations and theoretical fits to the CMB power spectrum.   
 Here we report on calculations of not only the numerical 
 CMB power spectrum from the PMF, 
 but also the correlations between
 the CMB power spectrum from the PMF
  and the primary curvature perturbations.
We then deduce a precise estimate of the PMF effect on all
 modes of perturbations.  
 We find that the PMF affects not only the CMB TT and TE modes on small
 angular scales, but also on large angular scales. 
 The introduction of a PMF  leads to a better fit to the CMB power spectrum for the higher multipoles, and the fit at lowest multipoles can be used to constrain the correlation of the PMF with the density fluctuations for large negative values of the spectral index.
 Our prediction for the BB mode for a PMF  average field strength $|B_\lambda| =4.0$ nG 
 is consistent with the upper limit on the BB mode deduced from the latest CMB
 observations.
 We find that the BB mode is dominated by the vector mode of the PMF for higher multipoles.
 We also show that by fitting the complete power spectrum one can break 
  the degeneracy between the  PMF amplitude and its  power spectral
 index. 
\end{abstract}

\pacs{98.62.En,98.70.Vc}
\keywords{Primordial magnetic field, CMB}
\maketitle
\section{Introduction}
Magnetic fields in clusters of galaxies have been observed 
\cite{Kronberg:1992pp,Wolfe:1992ab, Clarke:2000bz,Xu:2005rb}
 with a strength of $0.1-1.0~\mu$ G. 
The existence of a  primordial magnetic field (PMF) of order  1 nG  whose field lines collapse as structure forms is one possible explanation for such 
magnetic fields in galactic clusters.  The origin and detection of the PMF is, hence, a subject of considerable interest in modern cosmology. 
Moreover, the PMF could influence a variety of phenomena in the early universe
\cite{Grasso:2000wj} such as the cosmic microwave background (CMB)
\cite{
Challinor:2005ye,
Dolgov:2005ti,
Gopal:2005sg,
Kahniashvili:2005xe,
Kosowsky:2004zh,
Lewis:2004ef,
Mack:2001gc,
Subramanian:1998fn,
Subramanian:2002nh,
Yamazaki:2005yd,
Yamazaki:2004vq,
Yamazaki:2006bq,
Yamazaki:2006ah}, 
or the matter density field
\cite{
Giovannini:2004aw,
Tashiro:2005hc,
Tsagas:1999ft,
Yamazaki:2006mi}.

Temperature and polarization anisotropies in the CMB provide very
precise information on the physical processes in operation during the
early universe (WMAP \cite{Spergel:2006hy,Hinshaw:2006ia,Page:2006hz}, ACBAR
\cite{Kuo:2006ya}, CBI \cite{Readhead:2004gy,Sievers:2005gj}, DASI
\cite{Leitch:2004gd}, BOOMERANG \cite{Jones:2005yb}, and VSA
\cite{Dickinson:2004yr}).  The CMB power spectrum from ACBAR and CBI,
has indicated a potential discrepancy between these observations at
higher multipoles $\ell \ge 2000$ and the best-fit cosmological model to
the WMAP power spectrum.
A straightforward extension of the fit \cite{Spergel:2006hy} to the WMAP data
predicts a rapidly declining power spectrum  in the large multipole
range due to the finite thickness of the photon last scattering
surface and the Silk damping effect.  The  ACBAR and CBI
experiments, however,  indicate continued power up to $\ell \sim 4000$.
This discrepancy is difficult to account for by a simple retuning of
cosmological parameters or by the Sunyev-Zeldovich effect \cite{Spergel:2006hy,Komatsu02,Bond05}. 
Among other possible explanations, an 
inhomogeneous cosmological magnetic field
generated before the CMB last-scattering epoch provides a plausible
mechanism \cite{Bamba:2004cu} to produce excess power at high multipoles. Such a field
excites an Alfven-wave mode in the primordial baryon-photon plasma  and induces small rotational velocity  perturbations. Since this mode can survive on scales below those at which Silk
damping occurs during recombination
\cite{Jedamzik:1996wp,Subramanian:1998fn}, it could be a new source of
the CMB anisotropies on small angular scales. 
The present  work, therefore,  is an attempt to more precisely study the evolution of cosmological
perturbations with a PMF.

Previous work \cite{
Durrer:1999bk,
Seshadri:2000ky,
Campanelli:2004pm,
Challinor:2005ye,
Dolgov:2005ti,
Gopal:2005sg,
Kahniashvili:2005xe,
Kosowsky:2004zh,
Lewis:2004ef,
Mack:2001gc,
Subramanian:1998fn,
Subramanian:2002nh,
Yamazaki:2005yd,
Yamazaki:2004vq,
Yamazaki:2006bq,
Yamazaki:2006ah}
has shown that one can obtain information about the PMF
from the CMB temperature anisotropies and polarization.
However, in those works  attention was only given to a subset of the modes of the
CMB anisotropies.
In the present work, therefore, we  study the comprehensive effect of the PMF on all modes
of the CMB perturbations. 
Furthermore, in order to clarify  the role of the PMF in the CMB, 
we take into consideration the possible correlation between the CMB
fluctuations induced by the PMF and those due to primordial curvature and
tensor perturbations.  Also,    
we numerically
evaluate 
the CMB power spectrum from the stochastic PMF
 and thereby avoid recourse to analytic approximations.

In this article, we use adiabatic initial conditions for the
 evolution of primary density perturbations and consider isocurvature
 (isothermal) initial conditions when estimating effects on the CMB anisotropy induced by
 the PMF \cite{Lewis:2004ef,Giovannini:2006gz}.
Throughout this article we fix the best fit cosmological parameters of
the $\Lambda$CDM + Tensor model as follows \cite{Spergel:2006hy}: 
$h=0.792$, $\Omega_bh^2=0.02336$, $\Omega_m h^2=0.1189$, $n_S=0.987$,
$r=0.55$, $n_T=-r/8= -0.069$, and $\tau_c=0.091$ in flat universe models,
where $h$ denotes the Hubble parameter in units of 100 km s$^{-1}$Mpc$^{-1}$, $\Omega_b$
and $\Omega_m$ are the baryon and cold dark matter densities in units of the
critical density, $n_S$ is the spectral index of the primordial
scalar fluctuations, $r$ the ratio of the amplitude of the tensor fluctuations to
the scalar potential fluctuations, $n_T$ is the spectral index of the primordial
 tensor fluctuations, and $\tau_c$ is the optical depth for
Compton scattering. 
\section{Primordial Magnetic Field}
Before recombination, Thomson scattering between photons and
electrons along with Coulomb interactions between electrons and
baryons were sufficiently rapid that the photon-baryon system
behaved as a single tightly coupled fluid.
Since the trajectory of plasma particles is bent by Lorentz forces in a
magnetic field, photons are indirectly influenced by the magnetic field
through Thomson scattering. 
Let us consider the PMF created at some moment during the radiation-dominated epoch. 
The energy density of the magnetic field can be treated as a first order
perturbation upon a flat Friedmann-Robertson-Walker (FRW) background
metric.  
In the linear approximation, the magnetic field evolves as a stiff
source.
Therefore, we can discard all back reactions from the magnetohydrodynamic (MHD) fluid onto the field itself.
\subsection{Power Spectrum from the PMF}
To derive the power spectrum from the PMF we begin with 
the electromagnetic tensor in the usual form
\begin{eqnarray}
{F^\alpha}_\beta=
\left(
    \begin{array}{cccc}
      0    &  E_1  &  E_2  &  E_3 \\
      E_1  &  0    & -B_3  &  B_2 \\
      E_2  &  B_3  &  0    & -B_1 \\
      E_3  & -B_2  &  B_1  &  0  \\
    \end{array}
\right)~~,
\label{eq_emf_tensor}
\end{eqnarray}
where $E_i$ and $B_i$ are the electric and magnetic fields.
[Here we use natural units, i.e.~$c = \hbar = 1$.] 
The energy momentum tensor for electromagnetism is 
\begin{eqnarray}
{T^{\alpha\beta}}_{[\mathrm{EM}]}=\frac{1}{4\pi}\left(F^{\alpha\gamma}F^\beta_\gamma
-\frac{1}{4}g^{\alpha\beta}F_{\gamma\delta}F^{\gamma\delta}\right)\label{eq_ememtensor}.
\end{eqnarray}
The Maxwell stress tensor, $\sigma^{i k}$, is derived from the space-space components of the electromagnetic energy momentum tensor,  
\begin{eqnarray}
-{T^{ik}}_{[\mathrm{EM}]}=
\sigma^{ik}=  \nonumber \\
\frac{1}{a^2}\frac{1}{4\pi}\left\{E^i E^k+B^i B^k
- \frac{1}{2}\delta^{ik}(E^2+B^2)\right\}\label{eq_MST1}.
\end{eqnarray}

As mentioned above, within the linear approximation \cite{Durrer:1999bk} we can discard the MHD back reaction onto the field itself \cite{Brandenburg:1996sa}.
The conductivity of the primordial plasma is very large, and is "frozen-in" \cite{Ahonen:1997wh, Mack:2001gc}.
 This is a very good approximation during  the epochs of interest here.
 Furthermore, we can neglect the electric field, i.e.~$E\sim 0$, and can 
 decouple the time evolution of the magnetic field from its spatial dependence, i.e. $\mathbf{B}(\tau,\mathbf{x}) = \mathbf{B}(\mathbf{x})/a^2$ for very large scales \cite{Brandenburg:1996sa}. In this way we obtain the following equations,
\begin{eqnarray}
{T^{00}}_{[\mathrm{EM}]}=\frac{B^2}{8\pi a^6} \label{eq_MST_00}~~, \\
{T^{i0}}_{[\mathrm{EM}]}={T^{0k}}_{[\mathrm{EM}]}=0 \label{eq_MST_0s} ~~,\\
-{T^{ik}}_{[\mathrm{EM}]}=\sigma^{ik}=\frac{1}{8\pi a^6}(2B^i B^k -
\delta^{ik}B^2)~~.
\label{eq_MST_ss}
\end{eqnarray}
We assume that the PMF $\mbi{B}_0$ is statistically
homogeneous, isotropic and random.  
For such a magnetic field, the fluctuation power spectrum can be taken as a
power-law $S(k)=<B(k)B^\ast(k)> \propto k^{n_\mathrm{B}} $ \cite{Mack:2001gc} where $n_\mathrm{B}$ is the power-law
spectral index of the PMF.
The index $n_\mathrm{B}$ can be either negative or positive depending upon the
physical processes of magnetic field creation.
From Ref.~\cite{Mack:2001gc}, a two-point correlation function for the PMF can be  defined by
\begin{eqnarray}
\left\langle B^{i}(\mbi{k}) {B^{j}}^*(\mbi{k}')\right\rangle 
	&=&	\frac{(2\pi)^{n_\mathrm{B}+8}}{2k_\lambda^{n_\mathrm{B}+3}}
		\frac{B^2_{\lambda}}{\Gamma\left(\frac{n_\mathrm{B}+3}{2}\right)}
		k^{n_\mathrm{B}}P^{ij}(k)\delta(\mbi{k}-\mbi{k}'), 
		\ \ k < k_C~~,
		\label{two_point1} 
\end{eqnarray}
where
\begin{eqnarray}
P^{ij}(k)&=&
	\delta^{ij}-\frac{k{}^{i}k{}^{j}}{k{}^2}~~.
	\label{project_tensor}
\end{eqnarray}
 Here, $B_\lambda$ is the magnetic comoving mean-field amplitude obtained by smoothing over a Gaussian sphere of comoving radius $\lambda$,
and $k_\lambda = 2\pi/\lambda$ ($\lambda=1$ Mpc in this paper).
The cutoff wave number $k_C$ in the magnetic power
 spectrum is defined by \cite{Subramanian:1997gi},
\begin{eqnarray}
k_C^{-5-n_\mathrm{B}}(\tau)=
\left\{
		\begin{array}{rl}
			\frac{B^2_\lambda k_\lambda^{-n_\mathrm{B}-3}}{4\pi(\rho+p)}
			\int^{\tau}_{0}d\tau' 
			\frac{l_{\gamma}}{a},
			& \tau < \tau_\mathrm{dec} \\
			k_C^{-5-n_\mathrm{B}}(\tau_\mathrm{dec}), & \tau > \tau_\mathrm{dec},
		\end{array}
\right.
	\label{eq:CutOff_F}
\end{eqnarray}
where $l_\gamma$ is the mean free path of photons, and $\tau_\mathrm{dec}$ is the conformal time of the decoupling of photons from baryons.   
\subsection{Scalar Mode}
We obtain the power spectrum of the PMF energy density $|E_{\mathrm{[EM:S]}}(\mbi{k},\tau)|^2\delta(\mbi{k}-\mbi{k}')$ and the Lorenz force for the scalar modes $|\Pi_{\mathrm{[EM:S]}}(\mbi{k},\tau)|^2\delta(\mbi{k}-\mbi{k}')$ according to the following relations,
\begin{eqnarray}
|E_{\mathrm{[EM:S]}}(\mbi{k},\tau)|^2\delta(\mbi{k}-\mbi{k}')
=
\frac{1}{(2\pi)^3}
\left\langle
	T(\mbi{k},\tau)_{\mathrm{[EM:S1]}}T^*(\mbi{k}',\tau)_{\mathrm{[EM:S1]}}
\right\rangle~~,
\label{ED_Source}
\end{eqnarray}
and
\begin{eqnarray}
|\Pi_{\mathrm{[EM:S]}}(\mbi{k},\tau)|^2\delta(\mbi{k}-\mbi{k}')
=&&\frac{1}{(2\pi)^3}
	\left\langle
	\left(
		T(\mbi{k},\tau)_{\mathrm{[EM:S1]}}
		-T(\mbi{k},\tau)_{\mathrm{[EM:S2]}}
	\right)
	\right.
\nonumber\\
&&\times
	\left.
	\left(
		T^*(\mbi{k}',\tau)_{\mathrm{[EM:S1]}}
		-T^*(\mbi{k}',\tau)_{\mathrm{[EM:S2]}}
	\right)
	\right\rangle~~.
\nonumber\\
\label{LF_Souce}
\end{eqnarray}
In the case of a power law stochastic magnetic field, an explicit expression to evaluate the ensemble averages  for 
the above spectra 
is given by 
\begin{eqnarray}
\lefteqn{
	\langle
		T(\mbi{k},\tau)_{[\mathrm{EM:S1}]}
		T^*(\mbi{p},\tau)_{[\mathrm{EM:S1}]}
	\rangle
=	
	\frac{1}{2^4(2\pi)^8 a^{8}}
	\left\{
		\frac{(2\pi)^{n_\mathrm{B}+8}}{2k_\lambda^{n_\mathrm{B}+3}}
		\frac{B^2_{\lambda}}{\Gamma\left(\frac{n_\mathrm{B}+3}{2}\right)}
	\right\}^2
}\hspace{1cm}\nonumber\\
&&\times
	\int d^3k'
	k'{}^{n_\mathrm{B}}|\mbi{k}-\mbi{k}'|^{n_\mathrm{B}}
	\left\{
		1+\frac{\{\mbi{k}'\cdot(\mbi{k}-\mbi{k}')\}^2}
			   {k'{}^2|\mbi{k}-\mbi{k}'|^2}
	\right\}
	\delta(\mbi{k}-\mbi{p})~~.
\nonumber\\
\label{eq:S1S1_fnck}
\end{eqnarray}
Equivalently, in terms of an angular integration one can write,
\begin{eqnarray}
\lefteqn{
	\langle
		T(\mbi{k},\tau)_{[\mathrm{EM:S1}]}
		T(\mbi{p},\tau)^*_{[\mathrm{EM:S1}]}
	\rangle
=
\frac{1}{2^3(2\pi)^7 a^{8}}
\left\{\frac{(2\pi)^{n_\mathrm{B}+8}}{2k_\lambda^{n_\mathrm{B}+3}}\frac{B^2_{\lambda}}{\Gamma\left(\frac{n_\mathrm{B}+3}{2}\right)}\right\}^2
}\hspace{1cm}\nonumber\\
&&\times
\int dk'k'{}^{n_\mathrm{B}+2}\int^{1}_{-1} d\mathcal{C}
|\mbi{k}-\mbi{k}'|^{n_\mathrm{B}-2}
\left\{
(1+\mathcal{C}^2)k^2-4kk'\mathcal{C}+2k'{}^2
\right\}
\delta(\mbi{k}-\mbi{p})~~,
\nonumber\\  
&& 
\label{eq:S1S1_fncC}
\end{eqnarray}
where we define
\begin{eqnarray}
	\mathcal{C}
		\equiv  \cos{c}
		= \hat{\mbi{k}}\cdot\hat{\mbi{k}}'
		= \frac{\mbi{k}'\cdot\mbi{k}}{k'k}.\label{define_C}
\end{eqnarray}
In almost all previous work  the sum of the terms in brackets  which include
$\mathcal{C}$ in the $k$ integral of Eq.~(\ref{eq:S1S1_fncC}) have been set to unity. 
In this paper, however, we calculate Eq.~(\ref{eq:S1S1_fncC}) explicitly  by
integrating all of the terms.  In this way we obtain the following expression,
\begin{eqnarray}
\lefteqn{
	\langle
		T(\mbi{k},\tau)_{[\mathrm{EM:S1}]}
		T^*(\mbi{k},\tau)_{[\mathrm{EM:S1}]}
	\rangle
	=
	\frac{1}{8\pi a^{8}}
	\left\{
		\frac{(2\pi)^{n_\mathrm{B}+5}}{2k_\lambda^{n_\mathrm{B}+3}}
		\frac{B^2_{\lambda}}{\Gamma\left(\frac{n_\mathrm{B}+3}{2}\right)}
	\right\}^2
}\hspace{1cm}
 \nonumber \\
&&\times	
	\int dk'k'{}^{n_\mathrm{B}+2}
	\left[
	\frac{n_\mathrm{B}^2+4n_\mathrm{B}+1}{kk'n_\mathrm{B}(n_\mathrm{B}+2)(n_\mathrm{B}+4)}
	\left\{
		(k+k')^{n_\mathrm{B}+2}
		-|k-k'|^{n_\mathrm{B}+2}
	\right\}
	\right.
\nonumber\\
&&-
	\frac{1}{k'{}^2n_\mathrm{B}(n_\mathrm{B}+4)}
	\left\{
		|k-k'|^{n_\mathrm{B}+2}
		+|k+k'|^{n_\mathrm{B}+2}
\right\}
\nonumber\\
&&+
	\left.
	\frac{k}{k'{}^3n_\mathrm{B}(n_\mathrm{B}+2)(n_\mathrm{B}+4)}
	\left\{
		(k+k')^{n_\mathrm{B}+2}
		-|k-k'|^{n_\mathrm{B}+2}
	\right\}
	\right].
\label{eq:T1T1}
\end{eqnarray}
A similar calculation gives the power spectrum of the PMF tension and 
the power spectrum of the correlation between pressure and tension
 as follows, 
\begin{eqnarray}
\lefteqn{
	\left\langle 
		T_{[\mathrm{EM:S2}]}(\mbi{k})
		T^*_{[\mathrm{EM:S2}]}(\mbi{k})
	\right\rangle = 
	\frac{1}{2\pi a^{8}}
	\left\{
		\frac{(2\pi)^{n_\mathrm{B}+5}}{2k_\lambda^{n_\mathrm{B}+3}}
		\frac{B^2_{\lambda}}{\Gamma\left(\frac{n_\mathrm{B}+3}{2}\right)}
	\right\}^2
}\hspace{1cm}
	\nonumber\\ 
&&\times
	\int dk'
	k'{}^{n_\mathrm{B}+4}
	\frac{4}{(kk')^3n_\mathrm{B}(n_\mathrm{B}+2)(n_\mathrm{B}+4)}
	\left[\frac{}{}
		\left\{
			(k+k')^{n_\mathrm{B}+4}
			-|k-k'|^{n_\mathrm{B}+4}
		\right\}
	\right.
\nonumber\\
&&-
		\frac{3}{(kk')(n_\mathrm{B}+6)}
		\left\{
			|k-k'|^{n_\mathrm{B}+6}
			+(k+k')^{n_\mathrm{B}+6}
		\right\}
\nonumber\\
&&+
	\left.
		\frac{3}{(kk')^2(n_\mathrm{B}+6)(n_\mathrm{B}+8)}
		\left\{
			(k+k')^{n_\mathrm{B}+8}
			-|k-k'|^{n_\mathrm{B}+8}
		\right\}
	\right]~~,
\label{eq:T2T2} 
\end{eqnarray}
and
\begin{eqnarray}
\lefteqn{
	\langle
		T_{[\mathrm{EM:S1}]}(\mbi{k})
		T^*_{[\mathrm{EM:S2}]}(\mbi{k})
	\rangle
	+
	\langle
		T_{[\mathrm{EM:S2}]}(\mbi{k})
		T^*_{[\mathrm{EM:S1}]}(\mbi{k})
	\rangle
}\hspace{1cm}
\nonumber\\
&=&\frac{1}{2\pi a^{8}}
	\left\{
		\frac{(2\pi)^{n_\mathrm{B}+5}}{2k_\lambda^{n_\mathrm{B}+3}}
		\frac{B^2_{\lambda}}{\Gamma\left(\frac{n_\mathrm{B}+3}{2}\right)}
	\right\}^2
\nonumber\\
&&\times
    \int dk'k'{}^{n_\mathrm{B}+3}
\left[	
    \frac{1}{(kk')^2n_\mathrm{B}(n_\mathrm{B}+2)}
	\left\{
		(k+k')^{n_\mathrm{B}+3}
	   -|k-k'|^{n_\mathrm{B}+3}
	\right\}
\right.
\nonumber\\ 
&& 	-
	\frac{3}{k^2k'{}^3n_\mathrm{B}(n_\mathrm{B}+2)(n_\mathrm{B}+4)}
	\left\{
		|k-k'|^{n_\mathrm{B}+4}
		+(k+k')^{n_\mathrm{B}+4}
	\right\}
\nonumber\\ 
&& 	-
	\frac{1}{k^3k'{}^2n_\mathrm{B}(n_\mathrm{B}+2)(n_\mathrm{B}+4)}
	\left\{
		(k+k')^{n_\mathrm{B}+4}
	   -|k-k'|^{n_\mathrm{B}+4}
	\right\}
\nonumber\\ 
&&
\left.
	+\frac{3}{k^3k'{}^4n_\mathrm{B}(n_\mathrm{B}+2)(n_\mathrm{B}+4)(n_\mathrm{B}+6)}
	\left\{
		(k+k')^{n_\mathrm{B}+6}
	   -|k-k'|^{n_\mathrm{B}+6}
	\right\}
\right]~~.
\nonumber\\
\label{eq:T1T2}
\end{eqnarray}
\subsection{Vector Mode}
We obtain the power spectrum of the PMF Lorenz force for the vector mode 
from the following 
\begin{eqnarray}
|\Pi_\mathrm{[EM:V]}(\mbi{k})|^2\delta(\mbi{k}-\mbi{k}')
=
\frac{1}{2(2\pi)^3}
\left\langle
	T_i{}_\mathrm{[EM:V]}(\mbi{k})
	T_i^*{}_\mathrm{[EM:V]}(\mbi{k}')
\right\rangle ~~.
\end{eqnarray}
For the case of a power law stochastic magnetic field, an explicit expression 
for the ensemble average 
is given by 
\begin{eqnarray}
&&
\left\langle 
	T^{i}(\mbi{k})_\mathrm{[EM:V]}{T_{i}}^*(\mbi{k})_\mathrm{[EM:V]}
\right\rangle
\nonumber\\
&&=
	\frac{1}{4(2\pi)^7a^{8}}
		\left\{
			\frac{(2\pi)^{n_\mathrm{B}+8}}{2k_\lambda^{n_\mathrm{B}+3}}
			\frac{B^2_{\lambda}}{\Gamma\left(\frac{n_\mathrm{B}+3}{2}\right)}
		\right\}^2
	\int dk'
	\int^1_{-1} d\mathcal{C}
			k'{}^{n_\mathrm{B}+2}(\mbi{k}-\mbi{k}')^{n_\mathrm{B}-2}\nonumber\\
&&
  (1-\mathcal{C}^2)
  \left\{
    2k^2-5\mathcal{C}kk'
    +(2\mathcal{C}^2+1)k'{}^2
  \right\}.
\nonumber\\
\label{eq:<T_EM_V2><T_EM_V2>_IV}
\end{eqnarray}
Integrating Eq.(\ref{eq:<T_EM_V2><T_EM_V2>_IV}) over $\mathcal{C}$,
after a lengthy calculation, we  obtain the following expression,
\begin{eqnarray}
\lefteqn{
\left\langle 
	T^{i}(\mbi{k})_\mathrm{[EM:V]}{T_{i}}^*(\mbi{k})_\mathrm{[EM:V]}
\right\rangle
=
	\frac{1}{2(2\pi)^7 a^{8}}
		\left\{
			\frac{(2\pi)^{n_\mathrm{B}+5}}{2k_\lambda^{n_\mathrm{B}+3}}
			\frac{B^2_{\lambda}}{\Gamma\left(\frac{n_\mathrm{B}+3}{2}\right)}
		\right\}^2
	\int dk'
	k'{}^{n_\mathrm{B}+2}
}\hspace{1cm}
\nonumber\\
&\times&
    \frac{1}{n_\mathrm{B}(n_\mathrm{B}+2)}
 \left[
    \frac{1}{k^2k'{}^2}
	\left\{
		(2k-3k')(k-k')
		|k-k'|^{n_\mathrm{B}+2}
	\right.
\right.
\nonumber\\
&&
	\left.
		+(2k+3k')(k+k')
  		(k+k')^{n_\mathrm{B}+2}
  	\right\}
\nonumber\\
&&
-
  \frac{1}{k^3k'{}^3(n_\mathrm{B}+4)}
\left\{
-(11k'{}^2-15kk'+2k^2)
  |k-k'|^{n_\mathrm{B}+4}
	\right.
\nonumber\\
&&
	\left.
+(11k'{}^2+15kk'+2k^2)
  (k+k')^{n_\mathrm{B}+4}
\right\}
\nonumber\\
&&
-
  \frac{1}{k^4k'{}^4(n_\mathrm{B}+4)(n_\mathrm{B}+6)}
\left\{
(-24k'{}^2+15kk')
  |k-k'|^{n_\mathrm{B}+6}
	\right.
\nonumber\\
&&
	\left.
-(24k'{}^2+15kk')
  (k+k')^{n_\mathrm{B}+6}
\right\}
\nonumber\\
&&
+
\left.
  \frac{24}{k^5k'{}^3(n_\mathrm{B}+4)(n_\mathrm{B}+6)(n_\mathrm{B}+8)}
\left\{
  |k-k'|^{n_\mathrm{B}+8}-(k+k')^{n_\mathrm{B}+8}
\right\}
\right].\nonumber\\
\label{eq:PMF_V_Source}
\end{eqnarray}
\subsection{Tensor Mode}
We obtain the power spectrum of the  PMF for the tensor mode 
according to the following
\begin{eqnarray}
	|\Pi_\mathrm{[EM:T]}|^2\delta (\mbi{k}-\mbi{k}')
	=
	\frac{1}{4(2\pi)^3}
	\left\langle
		T^{ij}(\mbi{k})_\mathrm{[EM:T]}
		{T_{ij}}^*(\mbi{k}')_\mathrm{[EM:T]}
	\right\rangle.
\end{eqnarray}
For the case of a power law stochastic magnetic field, we use the explicit expression for the ensemble average as given by\cite{Durrer:1999bk} 
\begin{eqnarray}
\lefteqn{
\left\langle 
	T^{ij}(k)_\mathrm{[EM:T]}{T_{ij}}^*(k)_\mathrm{[EM:T]}
\right\rangle
 = 
	\frac{1}{4(2\pi)^7 a^8}
	\left\{
		\frac{(2\pi)^{n_\mathrm{B}+8}}{2k_\lambda^{n_\mathrm{B}+3}}
		\frac{B^2_{\lambda}}{\Gamma\left(\frac{n_\mathrm{B}+3}{2}\right)}
	\right\}^2
}\hspace{1cm}\nonumber\\
&&\times
	\int dk'
		k'{}^{n_\mathrm{B}}|\mbi{k}-\mbi{k}'|^{n_\mathrm{B}}
		\{1+\mathcal{C}^2\}\left\{1+\frac{(k-\mathcal{C}k')^2}{|\mbi{k}-\mbi{k}'|^2}\right\}
\nonumber\\
\label{eq:<T_EM_T2><T_EM_T2>_III} 
\end{eqnarray}
Again integrating Eq.~(\ref{eq:<T_EM_T2><T_EM_T2>_III}) over $\mathcal{C}$,
after a lengthy calculation, we  obtain the following,
\begin{eqnarray}
\lefteqn{
	\left\langle
		T^{ij}(\mbi{k})_\mathrm{[EM:T]}
		{T_{ij}}^*(\mbi{k})_\mathrm{[EM:T]}
	\right\rangle
	=\frac{1}{(2\pi)^7 a^{8}}
		\left\{
			\frac{(2\pi)^{n_\mathrm{B}+8}}{2k_\lambda^{n_\mathrm{B}+3}}
			\frac{B^2_{\lambda}}{\Gamma\left(\frac{n_\mathrm{B}+3}{2}\right)}
		\right\}^2
	\int dk'k'{}^{n_\mathrm{B}+2}
}\hspace{1cm}
\nonumber\\
&& \times
	\left[
	\frac{1}{-kk'n_\mathrm{B}}
		\{
			|k-k'|^{n_\mathrm{B}+2}
			-|k+k'|^{n_\mathrm{B}+2}
		\}
	\right.
\nonumber\\
&& -
		\frac{1}{(kk')^2n_\mathrm{B}(n_\mathrm{B}+2)}
		\left\{
			(2k'{}^2+k^2-4kk')
			|k-k'|^{n_\mathrm{B}+2}
		\right.
\nonumber\\
&&		\left.
			+
			(2k'{}^2+k^2+4kk')
			|k+k'|^{n_\mathrm{B}+2}
		\right\}
\nonumber\\
&&
-
	\frac{1}{(kk')^3n_\mathrm{B}(n_\mathrm{B}+2)(n_\mathrm{B}+4)}
	\left\{
		(4k'{}^2+k^2-6kk')
		|k-k'|^{n_\mathrm{B}+4}
		\right.
\nonumber\\
&&		\left.
		-(4k'{}^2+k^2+6kk')
		|k+k'|^{n_\mathrm{B}+4}
	\right\}
\nonumber\\
&&
-
	\frac{6}{(kk')^4n_\mathrm{B}(n_\mathrm{B}+2)(n_\mathrm{B}+4)(n_\mathrm{B}+6)}
	\left\{
			(k'{}^2-kk')|k-k'|^{n_\mathrm{B}+6}
		\right.
\nonumber\\
&&		\left.
			+
			(k'{}^2+kk')|k+k'|^{n_\mathrm{B}+6}
	\right\}
\nonumber\\
&&
-
		\frac{6k'{}^2}{(kk')^5n_\mathrm{B}(n_\mathrm{B}+2)(n_\mathrm{B}+4)(n_\mathrm{B}+6)(n_\mathrm{B}+8)}
		\left\{
			|k-k'|^{n_\mathrm{B}+8}
		\right.
\nonumber\\
&&	\left.
		\left.
			-
			|k+k'|^{n_\mathrm{B}+8}
		\right\}
	\right].
\nonumber\\
\label{eq:T_EM_T_TPCFC} 
\end{eqnarray}

These relations (Eqs.~\ref{ED_Source} - \ref{eq:T_EM_T_TPCFC}) constitute the various components of the PMF power spectrum evaluated in the present work.
\subsection{Numerical vs. Analytical}
\begin{figure*}[h]
\includegraphics[width=1.0\textwidth]{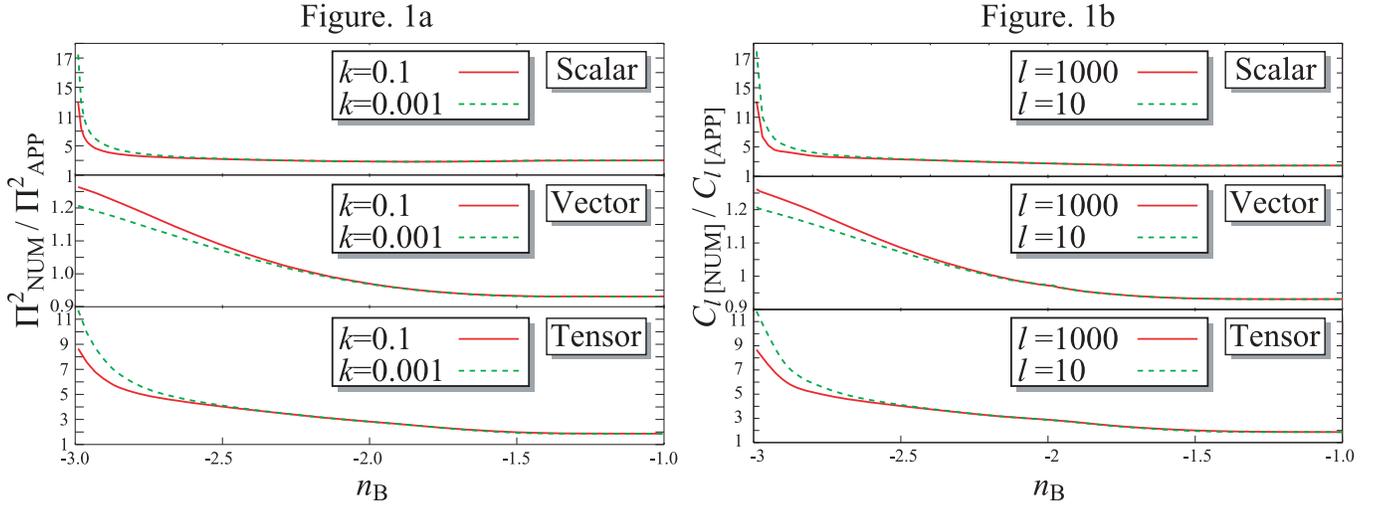}
\caption{\label{fig1}
Comparisons of approximated PMF power spectra for various modes  with  those evaluated
 numerically. 
The left panel (a) shows   the ratios of our numerical  estimations
 ($\Pi_\mathrm{NUM}^2$) to the previous approximations ($\Pi_\mathrm{APP}^2$)
 as a function of the power spectral index of the
 PMF($n_\mathrm{B}$). Bold (red) and dotted (green) curves are for $k=0.1$~Mpc$^{-1}$ and
 $0.001$~Mpc$^{-1}$, respectively.  The right panel (b) shows the the ratio of multipole coefficients $C_{\ell [NUM]}/ C_{\ell [APP]}$for 
 $\ell = 1000$ and $\ell = 10$ as labeled.}
\end{figure*}

In previous analyses
\cite{Caprini:2001nb,
Kahniashvili:2005xe,
Kosowsky:2004zh,
Lewis:2004ef,
Mack:2001gc,
Subramanian:2002nh} an analytic
approximation to the power spectrum of the PMF was utilized.
Specifically,
$
(k+k')^{n_\mathrm{B}+2}-|k-k'|^{n_\mathrm{B}+2}
 \sim
 2(n_\mathrm{B}+2)k^{n_\mathrm{B}+1}k'
$ 
for
$k' < k $, 
and
$ 2(n_\mathrm{B}+2)kk'{}^{n_\mathrm{B}+1} $
for
$k' > k $.
However, this is only a good approximation for $k' \ll k$ or
$k' \gg k$.  Outside of this range it can be a poor approximation since the neglected terms 
 are not small in  general.  In previous work
the source power spectrum from the PMF was obtained by integrating the approximate
equations over $0 < k' < k_C$, thus leading to
 smaller values than those evaluated
numerically for $n_\mathrm{B} <$ -2 as displayed in Figs.~\ref{fig1}a and~\ref{fig1}b.  
For example, when $n_\mathrm{B} = -2.9$,  (and $k < 0.1$) 
the relative errors in the squared 
power spectrum from the PMF ($\Pi^2_{NUM}/\Pi^2_{APP}$) can be 
as large as 400\%, 20\%, and 500\% for the scalar, vector
and tensor modes, respectively.  Figure 1b, shows the same result for the more familiar 
multipole coefficients, $C_\ell$, which are related to the various components of the power spectrum via an integration over wave number,
\begin{equation}
(2l+1)^2C^\mathrm{(X)}_\ell(PMF)  =
\frac{4}{\pi}
\int dk k^2
	[
	 \Theta^\mathrm{(X)}_\mathrm{PMF}(k)\Theta^\mathrm{(X)*}_\mathrm{PMF}(k)
	]~~.
\label{eq:CL}
\end{equation}
Here, $\Theta$ is the  photon moment, 
\begin{equation}
\Theta = T(k) \times \langle \Pi(k)^2 \rangle^{1/2} ~~, 
\label{moment}
\end{equation}
where $T(k)$ is the transfer function, and the index  $\mathrm{X} = \mathrm{S}$,  $\mathrm{V}$, or $\mathrm{T}$ denotes the scalar, vector, or tensor modes, respectively.  Lines are drawn on Figure 1b for $\ell = 10$ and $1000$, which roughly corresponds to the $k = 0.1$ Mpc$^{-1}$ and $0.001$ Mpc$^{-1}$  shown on Figure 1a. 

Figures 1a and 1b show that the deviations of the previous approximation from our numerical estimation increases as $n_\mathrm{B}$ decreases from $-1.5$ to  $-3.0$.  The deviations of the tensor and scalar modes in particular are as much as one order of magnitude near $n_\mathrm{B}=-3.0$.
We understand this for the following reasons:
Because integrands of all modes are dominated by values for range of $k' \le k$ for $n_\mathrm{B} < -1.5$, we can consider only this range. Since $k$ and $k'$ are smaller than unity, and for $n_\mathrm{B} < -2$, values of $|k-k'|^{n_\mathrm{B}+2}$ around $k\sim k'$ dominate the power spectrum of all modes (Eqs.~\ref{eq:T1T1} 
\ref{eq:T2T2}, \ref{eq:T1T2}, \ref{eq:T_EM_T_TPCFC}) except the vector mode (Eq.~\ref{eq:PMF_V_Source}), the deviations of the previous approximation from
 our numerical estimation  increase exponentially for $n_\mathrm{B} < -2$
 (Fig.~\ref{fig1}ab) \footnote{If the order of an integrand function $f(x) \propto x^{n}$ is more than -1, we can integrate $f(x$) numerically and analytically, for example, $\int (1/\sqrt{x}) dx \propto \sqrt{x}$}.
Since the order of the PMF source power spectra of all modes 
(\ref{eq:T1T1},
 \ref{eq:T2T2},
 \ref{eq:T1T2},
 \ref{eq:PMF_V_Source},
 \ref{eq:T_EM_T_TPCFC}) 
 is $2n_\mathrm{B}+3$,
 the values for $k'>k$ and $n_\mathrm{B} > -1.5$
 in integrands of all modes
 (\ref{eq:T1T1},
 \ref{eq:T2T2},
 \ref{eq:T1T2},
 \ref{eq:PMF_V_Source},
 \ref{eq:T_EM_T_TPCFC}) dominate the result of these integrations.
Therefore, if the integral range extends beyond $k$ for $n_\mathrm{B} > -1.5$, we can ignore the integrands of all modes for small enough $k'$ and the previous approximation is adequate for the PMF source power spectrum.
Thus, ratios of all modes are constant for $n_\mathrm{B} > -1.5$.
These results are almost the same as in Brown and Crittenden\cite{Brown:2005kr}
\footnote{
They, however, have only showed the result for a power spectral index $n_\mathrm{B} = 0$ and they have not analyzed nor explained the features of the approximation errors for $n_\mathrm{B} < -1.5$ in particular. 
Thus, we can not compare the our analysis of errors from the previous approximation method for $n_\mathrm{B} < -1.5$ with the result of Brown and Crittenden\cite{Brown:2005kr}.
}.
The constant ratios of all modes for $n_\mathrm{B} > -1.5$, however, 
 are not unity because the terms
including cosine factors [e.g.~ sum of terms within the bracket of Eq.~(\ref{eq:T1T1})], is not unity as was assumed in the previous approximation.

There is, however,  only a slight  deviation for the vector mode (cf.~the middle panels in Figs.~\ref{fig1}a and \ref{fig1}b).
This is because the power-law index of the $k-k'$ term in the PMF source power spectrum of the vector mode scales as (cf.~Eq.~\ref{eq:PMF_V_Source}) $n_\mathrm{B}+3$,  while the power-law index for the $(k-k')$ term for the scalar and tensor modes is $n_\mathrm{B} + 2$ in Eqs.~\ref{eq:T1T1} and 
\ref{eq:T_EM_T_TPCFC}.
Also, the $k+k'$ term slightly dominates the PMF source power spectrum of the vector mode ($\Pi_\mathrm{[EM:V]}(\mbi{k})$) for $n_\mathrm{B}<-1.5$.  Hence, unlike the scalar and tensor modes, there is only a slight deviation in the vector mode for $n_\mathrm{B}<-2$.
Additionally, since $k$ and $k'$ are smaller than unity and the power-law index of the
$(k+k')$ term, which dominates $\Pi_\mathrm{[EM:V]}(k=0.1)$, is positive,  
$\Pi_\mathrm{[EM:V]}(k=0.1) > \Pi_\mathrm{[EM:V]}(k=0.001)$.  This is in contrast to the scalar and tensor modes for which $\Pi_\mathrm{[EM:V]}(k=0.1) < \Pi_\mathrm{[EM:V]}(k=0.001)$.   The reason for this can be traced to the negative value of the power-law index $(n_\mathrm{B} + 2)$ for the scalar and tensor terms when $n_\mathrm{B} < - 2$.

  Also note on Figs.~\ref{fig1}a and \ref{fig1}b that even in the limit of $n_\mathrm{B} > -2$, the deviations to not asymptotically approach unity.  The reason for this is due to the inclusion of the correct $\mathcal{C}$ factors in the numerical angular integrals for Eqs.~(\ref{eq:S1S1_fncC}, \ref{eq:<T_EM_V2><T_EM_V2>_IV}, and \ref{eq:<T_EM_T2><T_EM_T2>_III}) which are not unity as is assumed in the analytic approximations. 

\section{Evolution Equations}
In this work, we have implemented a numerical method (\cite{Yamazaki:2006mi,Yamazaki:2007mm}) to evaluate the PMF source power spectrum.
Using this method, we are able to quantitatively  evolve the cut off scale and thereby reliably calculate the effects of the PMF on the observed CMB power spectrum.  We now summarize the essential evolution equations for each mode.
\subsection{Scalar Mode}
For the scalar mode we obtain the following equations in $k$-space
\cite{Padmanabhan:1993booka,
Ma:1995ey,
Hu:1997hp,
Hu:1997mn,
Dodelson:2003booka,
Giovannini:2006gz}:
\begin{eqnarray}
k^2\phi + 3H(\dot{\phi}+H\psi) &=& 4\pi G{a^2}
\left\{
	E_\mathrm{[EM:S]}(\mbi{k},\tau)-\delta\rho_\mathrm{tot}
\right\}\\
k^2(\phi-\psi) &=& 
-12\pi G{a^2}
\left\{
	Z_\mathrm{[EM:S]}(\mbi{k},\tau)
	-(\rho_\nu+P_\nu)\sigma_\nu
\right\}
\nonumber\\
 &=&
 -12\pi G{a^2}
 \left\{
 	\frac{1}{3}E_\mathrm{[EM:S]} 
 	(\mbi{k},\tau)+\Pi_\mathrm{[EM:S]}(\mbi{k},\tau)
	-(\rho_\nu+P_\nu)\sigma_\nu
 \right\}
\end{eqnarray}
\begin{eqnarray}
\dot{\delta}^\mathrm{(S)}&
			=&-(1+w)\left(v^\mathrm{(S)}+3\dot{\phi}\right)
			-3H\left(\frac{\delta p}{\delta\rho}-w\right)\delta^\mathrm{(S)}
\nonumber\\
&&			-\frac{3}{8\pi \rho}
			\left\{
				\dot{E}_{\mathrm{[EM:S]}}(\mbi{k},\tau)
				+6HE_{\mathrm{[EM:S]}}(\mbi{k},\tau)
			\right\}~,
			\label{eq:scalar_contiunuity1}\\
\dot{v}^\mathrm{(S)}&=&-H(1-3w)v^\mathrm{(S)}
			-\frac{\dot{w}}{1+w}v^\mathrm{(S)}
			+\frac{\delta p}{\delta\rho}
			\frac{k^2\delta^\mathrm{(S)}}{1+w}
			-k^2\sigma+k^2\psi
\nonumber\\
&&			
			+k^2 \frac{\Pi_{\mathrm{[EM:S]}}(\mbi{k},\tau)}
					 {4\pi \rho} ~,\label{eq:scalar_motion}
\end{eqnarray}
where $w \equiv p/\rho$.
Note that for the photon 
$\delta^\mathrm{(S)}_\gamma = 4\Theta^\mathrm{(S)}_0$, and 
$v^\mathrm{(S)}_\gamma=k\Theta^\mathrm{(S)}_1$.
Massless neutrinos obey Eqs.~(\ref{eq:scalar_contiunuity1}) and (\ref{eq:scalar_motion}) without the Thomson coupling term.
In the continuity and Euler relations (Eqs.~\ref{eq:scalar_contiunuity1} and \ref{eq:scalar_motion}) for the scalar mode, we can
just add the energy density and pressure of the PMF to the 
energy density and pressure of cosmic fluids, respectively. Since the baryon
fluid behaves like a nonrelativistic fluid during the epoch of interest, we
may neglect $w$ and $\delta P^\mathrm{(S)}_b/\delta\rho^\mathrm{(S)}_b$, except the acoustic term 
$c_sk^2\delta^\mathrm{(S)}_b$. Also, the shear stress of baryons is negligible \cite{Ma:1995ey}. 
Since we concentrate on scalar type
perturbations in this paper, we do not consider the magneto-rotational
instability from the shear stress of the PMF and baryon fluid
\cite{Chandrasekhar:1961ab}.

 From equations (\ref{eq:scalar_contiunuity1}) and
 (\ref{eq:scalar_motion}), we obtain the same form for  the evolution
 equations of photons and baryons as in previous
 work \cite{Padmanabhan:1993booka,Ma:1995ey,Hu:1997hp,Hu:1997mn,Dodelson:2003booka}, by considering the Compton interaction
 between baryons and photons,  
\begin{eqnarray}
k^2\phi + 3H(\dot{\phi}+H\psi) &=& 4\pi G{a^2}
\left\{
	E_\mathrm{[EM:S]}(\mbi{k},\tau)-\delta\rho_\mathrm{tot}
\right\} \label{eq:phi}\\
k^2(\phi-\psi) &=& 
-12\pi G{a^2}
\left\{
	Z_\mathrm{[EM:S]}(\mbi{k},\tau)
	-(\rho_\nu+P_\nu)\sigma_\nu
\right\}
\nonumber\\
 &=&
 -12\pi G{a^2}
 \left\{
 	\frac{1}{3}E_\mathrm{[EM:S]} 
 	(\mbi{k},\tau)+\Pi_\mathrm{[EM:S]}(\mbi{k},\tau)
	-(\rho_\nu+P_\nu)\sigma_\nu
 \right\}
 \label{eq:phi_psi}\\
\dot{\delta}^\mathrm{(S)}_\mathrm{CDM}
 	&=&
 		-v^\mathrm{(S)}_\mathrm{CDM}+3\dot{\phi}~,\label{eq:CDM_rho}\\
\dot{v}^\mathrm{(S)}_\mathrm{CDM}
 	&=&
        -\frac{\dot{a}}{a}v^\mathrm{(S)}_\mathrm{CDM}+k^2\psi~,\label{eq:CDM_v}\\
\dot{\delta}^\mathrm{(S)}_{\gamma}
 	&=&
 		-\frac{4}{3}v^\mathrm{(S)}_{\gamma}
 		+4\dot{\phi}~,\label{eq:photon_rho}\\
\dot{\delta}^\mathrm{(S)}_{\nu}
 	&=&
 		-\frac{4}{3}v^\mathrm{(S)}_{\nu}
 		+4\dot{\phi}~,\label{eq:photon_nu}\\
\dot{v}^\mathrm{(S)}_{\gamma}
	&=&
		k^2\left(\frac{1}{4}\delta^\mathrm{(S)}_{\gamma}-\sigma_{\gamma}\right)
		+an_e\sigma_T(v^\mathrm{(S)}_\mathrm{b}-v^\mathrm{(S)}_{\gamma})~+k^2\psi,
		\label{eq:photon_v} \\
\dot{v}^\mathrm{(S)}_{\nu}
	&=&
		k^2\left(\frac{1}{4}\delta^\mathrm{(S)}_{\nu}-\sigma_{\nu}\right)+k^2\psi,\label{eq:photon_nu} \\
\dot{\delta}^\mathrm{(S)}_\mathrm{b}
	&=&
		-v^\mathrm{(S)}_\mathrm{b}+3\dot{\phi}
 			\label{eq:baryon_rho}  \\
\dot{v}^\mathrm{(S)}_\mathrm{b}
	&=&
			-\frac{\dot{a}}{a}v^\mathrm{(S)}_\mathrm{b}
 			+c^2_sk^2\delta^\mathrm{(S)}_\mathrm{b}
 			+\frac{4\bar{\rho}_\gamma}{3\bar{\rho}_\mathrm{b}}
 			an_e\sigma_T(v^\mathrm{(S)}_{\gamma}-v^\mathrm{(S)}_\mathrm{b})+k^2\psi
\nonumber\\
&&			
			+\frac{3}{4}k^2\frac{\Pi_{\mathrm{[EM:S]}}(\mbi{k},\tau)}{R\rho_\gamma}~,
			\label{eq:baryon_v}
\end{eqnarray} 
where $R\equiv(3/4)(\rho_b/\rho_\gamma)$ is the inertial density ratio between baryons and photons, $n_e$ is the free electron density, $\sigma_T$ is the Thomson
scattering cross section, and $\sigma_{\gamma}$ of the second term on
the right hand side of equation (\ref{eq:photon_v}) is the
shear stress of the photons with the PMF. 
Since $n_\mathrm{B}\lesssim 0$ is favored by constraints from the gravitational wave 
background \cite{Caprini:2001nb} and the effect of the PMF is not influenced by the time evolution of
the cut off scale $k_C$ for this range of $n_\mathrm{B}$,
we approximately set $E_\mathrm{[EM:S]}\propto a^{-4}$ in the following analysis.
\subsection{Vector Mode}
The evolution of the vector potential $V(\tau, \mathbf{k})$ under the influence of a stochastic PMF can be written \cite{Hu:1997hp,Hu:1997mn} as
\begin{eqnarray}
\dot{V}+2\frac{\dot{a}}{a}V
=-\frac{16\pi a^2 G \Pi_\mathrm{[EM:V]} (\mathbf{k},\tau)}{k}
\nonumber \\
-8\pi Ga^2\frac{p_\gamma\pi_\gamma+p_\nu\pi_\nu}{k}~~,
\label{eq:vectorpoten}
\end{eqnarray}
where the dot denotes a conformal time derivative, while $p_i$ and $\pi_i$ are the pressure and the anisotropic stress of the photons ($i=\gamma$) and neutrinos ($i=\nu$).
 Here, we have omitted the vector anisotropic stress of the plasma which is negligible in general. 
In the absence of a magnetic source term,
the homogeneous solution of Eq.~(\ref{eq:vectorpoten}) behaves like $V\propto1/a^2$. 
We take $a\propto\tau$ during the radiation-dominated epoch.
The magnetic field, therefore, causes the vector perturbations to decay less
rapidly ($\propto 1/a$ instead of $1/a^2$) with the universal expansion.

Since the vector perturbations cannot generate density perturbations, we have $\delta^\mathrm{(V)}_\gamma=\delta^\mathrm{(V)}_b=0$, where $\delta^\mathrm{(V)}_\gamma$ and $\delta^\mathrm{(V)}_b$ are the perturbations of the photon and baryon energy densities, respectively. 

The magnetic field affects the photon-baryon fluid
dynamics via a Lorentz force
term in the baryon Euler equations. Following \cite{Hu:1997hp,Hu:1997mn}, the Euler equations for the neutrino, photon and baryon velocities, $v^\mathrm{(V)}_\nu$, 
$v^\mathrm{(V)}_\gamma$, and $v^\mathrm{(V)}_b$ are written as
\begin{eqnarray}
\dot{v}_{\nu}^\mathrm{(V)}-\dot{V}=-k\left(\frac{\sqrt{3}}{5}\Theta^\mathrm{(V)}_{\nu 2}\right)~~,
\label{eq:vneutrino1}\\
 \dot{v}_{\gamma}^\mathrm{(V)}
-\dot{V}
+\dot{\tau}_c
    (
	v^\mathrm{(V)}_{\gamma}-v^\mathrm{(V)}_{b}
	)
=-k\left(
			\frac{\sqrt{3}}{5}\Theta^\mathrm{(V)}_{\gamma 2}
   \right)~~,
\label{eq:vphoton1}\\
 \dot{v}_{b}^\mathrm{(V)}
-\dot{V}
+\frac{\dot{a}}{a}(v_{b}^\mathrm{(V)}-V)
-\frac{1}{R}\dot{\tau}_c(v^\mathrm{(V)}_{\gamma}-v^\mathrm{(V)}_{b})\nonumber \\
=\frac{3}{4}\frac{\Pi_\mathrm{[EM:V]}(\mathbf{k},\tau)}{R\rho_\gamma}~~.
\label{eq:vbaryon1}
\end{eqnarray}
For the photons 
$v^\mathrm{(V)}_\gamma=\Theta^\mathrm{(V)}_1$, while 
$\Theta^\mathrm{(V)}_{\nu 2}$ and $\Theta^\mathrm{(V)}_{\gamma 2}$ are quadrupole moments of the neutrino and photon angular distributions, respectively. These quantities are proportional to the anisotropic stress tensors. Equations (\ref{eq:vneutrino1})-(\ref{eq:vbaryon1}) denote the vector equations of motion for the cosmic fluid, which arise from the conservation of energy-momentum. 
\subsection{Tensor mode}
The Einstein equations tell us that the tensor mode $\mathcal{H}$  is governed
by \cite{Padmanabhan:1993booka,Hu:1997hp,Hu:1997mn,Dodelson:2003booka}
\begin{eqnarray}
\ddot{\mathcal{H}}
+2\frac{\dot{a}}{a}\dot{\mathcal{H}}
+k^2\mathcal{H}=
8\pi G a^2
\left( 
		\Pi_\mathrm{[EM:T]}+
		\Pi^\mathrm{(T)}_\nu
\right)
~~,
\end{eqnarray}
where $\Pi^\mathrm{(T)}_\nu$ is the anisotropic stress for neutrinos.
\subsection{Initial Conditions}
We need to specify the initial perturbations for solving the evolution equations derived in the previous sections.
We start solving the  evolution equations at early times when a given $k$ mode
is still outside the horizon, i.e.~the dimensionless parameter  $k\tau \ll 1$.
We consider only the radiation-dominated epoch since the numerical integration for all of the $k$ modes of interest will start in this era. 
Baryons and photons are tightly coupled at this early time.
The expansion rate is $H = \tau^{-1}$.
We then derive initial conditions for all of the modes utilizing  the method of
Refs.~\cite{Padmanabhan:1993booka,Ma:1995ey,Hu:1997hp,Hu:1997mn,Dodelson:2003booka,Lewis:2004ef,Kahniashvili:2007xe}.
\subsubsection{scalar mode initial condition}
We can assume that all density fields are  zero since initially, since
the PMF only affects the velocity field of ionized baryons,  i.e.~the  Lorenz force, and the density fields are not directly affected by the PMF.
In the radiation dominated era, photons and neutrinos are important in the RHS of Einstein equations. Equations for photons and neutrinos are
\begin{eqnarray}
\dot{\delta}^\mathrm{(S)}_{\gamma}
 	&=&
 		-\frac{4}{3}v^\mathrm{(S)}_{\gamma}
 		+4\dot{\phi}~,\label{eq:photon_rho_early}\\
\dot{\delta}^\mathrm{(S)}_{\nu}
 	&=&
 		-\frac{4}{3}v^\mathrm{(S)}_{\nu}
 		+4\dot{\phi}~,\label{eq:nu_rho_early}\\
\dot{v}^\mathrm{(S)}_{\gamma}
	&=&
		k^2\frac{1}{4}\delta^\mathrm{(S)}_{\gamma}
		+k^2\psi,
		\label{eq:photon_v_early} \\
\dot{v}^\mathrm{(S)}_{\nu}
	&=&
		k^2
		\left(
				 \frac{1}{4}\delta^\mathrm{(S)}_{\nu}
				-\sigma^\mathrm{(S)}_{\nu}
				+k^2\psi
		\right),
		\label{eq:nu_v_early}\\
\sigma^\mathrm{(S)}_{\nu}
	&=&
		\frac{4}{15}v^\mathrm{(S)}_\gamma
.\label{eq:nu_v_early}
\end{eqnarray}
Here we omitted higher multipole moments $\ell > 1$ for photons and $\ell > 2$ for neutrinos. 
At the lowest order in $k\tau$, initial conditions of Eqs.(\ref{eq:phi}-\ref{eq:baryon_v})
\begin{eqnarray}
\delta^\mathrm{(S)}_\gamma  = 
\delta^\mathrm{(S)}_\nu  =
\frac{4}{3}\delta^\mathrm{(S)}_b  =
\frac{4}{3}\delta^\mathrm{(S)}_\mathrm{CDM}  &=&
	R_\gamma R_\mathrm{B}
	+4R_\gamma
		\frac{4\sigma_\mathrm{B}+R_\nu R_\mathrm{B}}
			 {4R_\nu+15},\\
v^\mathrm{(S)}_\gamma = 
v^\mathrm{(S)}_b = 
v^\mathrm{(S)}_\mathrm{CDM} &=& 
-\frac{19}{4}
		\frac{4\sigma_\mathrm{B}+R_\nu R_\mathrm{B}}
			 {4R_\nu+15}
			 k^2\tau,\\
v^\mathrm{(S)}_\nu &=& 
-\frac{15}{4}
 \frac{R_\gamma}{R_\nu}
		\frac{4\sigma_\mathrm{B}+R_\nu R_\mathrm{B}}
			 {4R_\nu+15}
			 k^2\tau,\\
\sigma^\mathrm{(S)}_\nu &=& 
 -\frac{R_\gamma}{R_\nu}\sigma_\mathrm{B}
 +
 \frac{R_\gamma}{2R_\nu}
		\frac{4\sigma_\mathrm{B}+R_\nu R_\mathrm{B}}
			 {4R_\nu+15}
			 k^2\tau^2,\\
\psi = 
-2\phi &=& 
 -2R_\gamma
 \frac{R_\gamma}{2R_\nu}
		\frac{4\sigma_\mathrm{B}+R_\nu a R_\mathrm{B}}
			 {4R_\nu+15}.
\end{eqnarray}
where 
\begin{eqnarray}
\sigma_\mathrm{B} 
	&\equiv&
		-\frac{Z_\mathrm{[EM:S]}}{\rho_\gamma+P_\gamma} ,\nonumber\\
R_\gamma
	&\equiv&
		\frac{\rho_\gamma}{\rho_\gamma+\rho_\nu},\nonumber\\
R_\nu
	&\equiv&
		\frac{\rho_\nu}{\rho_\gamma+\rho_\nu},\nonumber\\
R_\mathrm{B}
	&\equiv&
		\frac{E_\mathrm{[EM:S]}}{\rho_\gamma}\nonumber
\end{eqnarray}
\subsubsection{vector mode initial condition}
At early times we neglect the vector anisotropic stress of the plasma, 
which is in general small.
Equation (\ref{eq:vectorpoten}) gives
\begin{eqnarray}
\dot{V} (\tau, \mathbf{k})+2\frac{\dot{a}}{a}V(\tau, \mathbf{k})
=-\frac{16\pi a^2 G \Pi_\mathrm{[EM:V]} (\tau, \mathbf{k})}{k}~~,
\label{eq:ini_vectorpoten}
\end{eqnarray}
which can be easily solved to obtain \cite{Mack:2001gc}
\begin{eqnarray}
V=-\frac{16\pi G a^2\Pi_\mathrm{[EM:V]} (\tau, \mathbf{k})}{k}\tau~~.
\label{eq:ini_vectorpoten_solv}
\end{eqnarray}
From Eqs.(\ref{eq:vphoton1}) and (\ref{eq:vbaryon1}), we obtain \cite{Mack:2001gc,Lewis:2004ef}
\begin{eqnarray}
{v}^\mathrm{(V)}_b=
	-\frac{3}{4}
	\frac{\Pi_\mathrm{[EM:V]}(\tau, \mathbf{k}) k\tau}{\rho_\gamma(1+R)}
	-\frac{16\pi G a^2\Pi_\mathrm{[EM:V]} (\tau, \mathbf{k})}{k}\tau~~.
\end{eqnarray}
\subsubsection{tensor mode initial condition}
From Ref.~\cite{Lewis:2004ef},
we obtain the initial condition of the tensor mode as follows
\begin{eqnarray}
\mathcal{H}^{(0)}=3R_\gamma
\ln
\left(
	\frac{\tau^{\nu}}{\tau_\mathrm{PMF}}
\right)\\
\mathcal{H} = 
\left\{
	\mathcal{H}^{(0)}
		\left(
			1
			-\frac{5}{8R_\nu + 30}
		\right)
		+
			\frac{15}{28}
			\frac{R_\gamma }
			     {4R_\nu + 15}
\right\}
\frac{\Pi_{\mathrm{[EM:T]}}}
{\rho_\gamma}
(k\tau)^2\\
\Pi^\mathrm{(T)}_\nu =
\left\{
\frac{4}{3}
\frac{\mathcal{H}^{(0)}k^2\tau^2}
     {4R_\nu+15}
-
\left(
 1
-\frac{15}{14}
 \frac{k^2\tau^2}
      {4R_\nu+15}
\right)
\frac{R_\gamma}{R_\nu}
\right\}
\frac{\rho_\nu}{\rho_\gamma}
\Pi_\mathrm{[EM:T]}~~,
\end{eqnarray}
where $\tau^{\nu}$ is the time of neutrino decoupling, and
$\tau_\mathrm{PMF}$ is the time of generating the PMF,
 $R_\gamma \equiv\Omega_\gamma/(\Omega_\nu+\Omega_\gamma)$,
and $R_\nu \equiv\Omega_\nu/(\Omega_\nu+\Omega_\gamma)$.
\section{Correlations in the Power Spectra}
Although possible origins of the PMF have been studied by many authors,
there is no consensus as to how the PMF correlates with the primordial density
fluctuations. 
Nonetheless, almost all previous works have  assumed that there is no
correlation between them \cite{Tashiro:2005hc}. 
In order to study the possible effects of a PMF in a more general manner, 
we here consider possible  correlations between the PMF and the primordial
density and tensor fluctuations. 
To do this we introduce a coefficient $s^\mathrm{(X)}$ to parameterize the correlation between 
the PMF source and the primary power spectrum
 \cite{Yamazaki:2006mi, Giovannini:2006gz}.  
The generalized multipole coefficients  $C^\mathrm{(X)}_l$ 
(with $\mathrm{X} = \mathrm{S}$ or $\mathrm{T}$ for scalar or tensor modes) then become,
\begin{eqnarray}
(2\ell + 1)^2C^\mathrm{(X)}_\ell &=&
\frac{4}{\pi}
\int dk k^2
	[
	 \Theta^\mathrm{(X)}_\mathrm{p}(k)\Theta^\mathrm{(X)*}_\mathrm{p}(k)
	+\Theta^\mathrm{(X)}_\mathrm{PMF}(k)\Theta^\mathrm{(X)*}_\mathrm{PMF}(k)
	\nonumber \\
	&&+s^\mathrm{(X)}
	\{
		\Theta^\mathrm{(X)}_\mathrm{p}(k) \Theta^\mathrm{(X)*}_\mathrm{PMF}(k)
		+\Theta^\mathrm{(X)}_\mathrm{PMF}(k) \Theta^\mathrm{(X)*}_\mathrm{p}(k)
	\}
	]
,\label{eq:col1}
\end{eqnarray}
where $\Theta$ is the  photon moment as defined in Eq.~\ref{moment}.
The last term is the correlation between the primary fluctuations and 
the power spectrum from the PMF.
Among the many possible sets of two correlation coefficients that satisfy 
$-1\le s^\mathrm{(X)} \le 1$, 
the two cases of $s^\mathrm{(S)}=s^\mathrm{(T)}=\pm 1$
 are expected to show the maximum absolute effects from the PMF.  These limits  represent the effective range of the resultant CMB anisotropies when a PMF is present.

\section{Results and Discussion}
We have explored effects of a PMF on the CMB for the allowed PMF
 parameters 
which were  deduced in our previous work \cite{Yamazaki:2006bq} (i.e.~$B_\lambda < 10~ \mathrm{nG~and~} n_\mathrm{B} < -2.4$) . 
The upper panel of Figure 2 illustrates the CMB temperature and polarization anisotropies from the PMF for scalar, vector and tensor modes for the case when $B_\lambda = 4.0$ nG and $n_\mathrm{B} = -2.9$ or $-2.5$ as labeled.
The scalar mode dominates for lower $\ell$ of the TT and TE modes
 as shown by Giovannini \cite{Giovannini:2006gz}.  
In particular, it is comparable in power to the primary TT mode for
 ($B_\lambda, n_\mathrm{B}) = (4.0 ~{\rm nG}, -2.9$). 
 We note that the curves with  ($B_\lambda, n_\mathrm{B}, s^\mathrm{(S)},s^\mathrm{(T)} ) = (4.0 ~{\rm nG}, -2.5, 1,1$)
 give the best fits to the observed power spectra in both the regions of high and low $\ell$.  This result  is complementary to and consistent with the nucleosynthesis constraints derived in \cite{Caprini:2001nb} as shown in \cite{Yamazaki:2007mm}.
 
For illustration, let us assume the same 
 scale-invariant power 
 spectrum for both the PMF and the primordial curvature perturbations.
In this case, the ratio of the density and velocity perturbations induced by
the primordial curvature perturbation to those by the PMF is
 proportional to $k^2$.
Therefore, the temperature anisotropies from the PMF are larger for
 lower $\ell$ compared with those from primordial curvature perturbations. 
Furthermore, the power of the CMB temperature anisotropies from the PMF
 for lower $\ell$ depends not only on $B_\lambda$ but also strongly on $n_\mathrm{B}$ [Panel (1a) of Fig.2].   Note, that the magnetic field which is continually sourcing fluctuations does not
 spoil the phase coherence (cf.~cosmic defects).
The basic behavior of acoustic oscillations is affected by the pressure of the fluid $kc_s\delta^\mathrm{(S)}_b$ and the potential $k\phi$. Since the pressure of the PMF is sufficiently less than the thermal fluid pressure, the PMF dose not affect  the phase coherence significantly.

The PMF also affects the CMB power spectrum on  small angular scales for two reasons.
First, the PMF energy density fluctuations depend only on the scale
factor $a$ and  can survive below Silk damping scale.
Therefore, the PMF continues to  source the fluctuations through the Lorentz force even below the
Silk damping scale.
Second, the vector mode from the PMF can be larger than
the scalar modes both from the PMF and primordial perturbations at small
scales.   
This is because, after horizon crossing, the latter cannot grow due to
the photon pressure leading to acoustic oscillations, while the former
can keep growing inside the cosmic horizon.
This means that, for higher $\ell$, the vector mode 
dominates the temperature anisotropies of the CMB over the scalar
and tensor modes from the PMF and the contribution from primary anisotropies. 
The integrated amplitude of the gravitational waves from the PMF can be
negligible after horizon crossing.  This is because the homogeneous solution
begins to oscillate inside the horizon and decay rapidly
\cite{
1979ZhPmR..30..719S,
1982PhLB..115..189R,
1985SvA....29..607P,
Pritchard:2004qp}. 
Consequently, gravity waves  only affect the anisotropy spectrum  on
scales larger than the horizon at recombination.  The tensor mode
from the PMF  therefore decreases at higher $\ell$. 
Thus, the vector mode of the CMB polarization from the PMF dominates for
higher $\ell$ [Panel (1d) of Fig.~2] \cite{Seshadri:2000ky,Lewis:2004ef}
\footnote{We do not consider the magnetic-field-related BB 
 polarization coming from the Faraday rotation effect
 \cite{Campanelli:2004pm,Kosowsky:2004zh}.}
In the primary spectrum for higher $\ell$
the absolute value of the EE mode from the PMF is relatively-small [Panel (1c) of Fig.~2] compared to the TT mode [Panel (1a) of Fig.~2].  Hence, even though the EE mode of the primary spectrum damps less than the TT mode, the EE mode remains much smaller than the TT mode at higher $\ell$ in the final power spectrum.
 For the TE mode [Panel (1b) of Fig.~2], the contribution from the PMF vector
and scalar modes can be comparable to that from the primordial curvature  perturbations
for higher $\ell$.   Except for a small dip region near $\ell = 40$, the scalar mode is always small compared to the primary spectrum.

Regarding the BB mode, we note that 
the  BB mode signal described in this paper is due to magnetic-field-induced CMB
 fluctuations (with the peak around $\ell \sim 2000$ as in \cite{Seshadri:2000ky,Lewis:2004ef}).  We do
 not include magnetic-field related BB 
 polarization coming from the Faraday rotation effect (with the peak around 
$\ell > 15000$) discussed in \cite{Campanelli:2004pm,Kosowsky:2004zh}. 
In our model, the BB mode from the PMF can dominate for $\ell \gtrsim 200$ if $B_{\rm
\lambda}\gtrsim 2.0$ nG.
 A potential problem in attempting to detect this signal on such
angular scales, therefore,  is the contamination from gravitational lensing
 which converts the dominant EE power into the BB mode
\cite{Pritchard:2004qp}. 
However, since we already know quite accurately what the spectrum of the lensing signal must be, we can subtract its power directly.
After removing the foreground effect, the BB mode from the PMF
effect dominates for higher $\ell$ even for PMF parameters allowed
by the CMB temperature constraint
\cite{Yamazaki:2004vq,Yamazaki:2006bq,Caprini:2001nb}.
Note, that the there is a change of scale between panels 1c and 1b.  The BB mode and EE modes are of comparable magnitude.

In Panel 2 of Fig. 2 we depict the CMB temperature and polarization anisotropies in the presence of a
 PMF  taking
into account the correlations.
Since we obtain the temperature and polarization anisotropies of
 the CMB with the PMF from isocurvature initial conditions, the phase
 of the CMB perturbation with the PMF is different by $\pi /2$ from those
 without the PMF (on the adiabatic initial condition). 
 The first, third, and odd numbered peaks of the scalar mode of the CMB perturbations rise for a positive correlation between the PMF and the energy density perturbations, while they are suppressed for a negative correlation.
These are compared in Panel 2 of Figure 2 with the observed power spectra
(WMAP \cite{Hinshaw:2006ia,Page:2006hz}, ACBAR \cite{Kuo:2006ya},
CBI \cite{Readhead:2004gy,Sievers:2005gj}, DASI \cite{Leitch:2004gd},
BOOMERANG \cite{Jones:2005yb}, and VSA \cite{Dickinson:2004yr}). 
Panels (2a) and (2b) of Fig.~2 show clearly that the power spectral index of the PMF, $n_\mathrm{B}$, is more effectively constrained from CMB observations for lower $\ell$ than those for higher $\ell$. The models with higher $n_\mathrm{B}$ give better fits to the observations than those with lower $n_\mathrm{B}$ for the lower $\ell$ 
regions of the TT and TE modes. 
Furthermore, there is no discrepancy at higher $\ell$ between observations and theories 
of the CMB polarization for models with a PMF 
[Panels (2c) and (2d) of Fig.2].
In  our previous work
 \cite{Yamazaki:2004vq,Yamazaki:2006bq} there was a problem from the strong degeneracy between $n_\mathrm{B}$ and
 $B_\lambda$.
This degeneracy, however,  is broken by the
different effects of the PMF on the CMB power spectrum for lower and higher $\ell$.

The scalar-mode CMB temperature-anisotropy power-spectrum shape
 ($\ell$-scaling for different $n_\mathrm{B}$ at large scales  - low  $\ell$) agrees with the results of
 \cite{Kahniashvili:2007xe}, while the E-polarization 
power spectrum shape does not follow semi-analytical estimate in that paper.
This difference is caused by the fact that we have included the effects of reionization \cite{Dodelson:2003booka, Lewis02} which were neglected in \cite{Kahniashvili:2007xe}.  
If the universe is re-ionized at $z_\mathrm{re}$, 
CMB photons are scattered by electrons and the polarization is generated again.
Since re-ionization  results in a new scattering surface  at relatively short distances from us at (and a relatively recent era), the viewing angle of the polarization from re-ionization becomes large.
Thus, reionization causes some power to shift to lower multipoles \cite{Kaplinghat:2002vt, Hu:2003gh}.  

We also point out that the observed power spectrum of temperature fluctuations in the CMB is likely to depend on frequency 
\cite{Kuo:2006ya}.  Such dependence is theoretically expected to
originate from foreground effects such as the Sunyaev-Zel'dovich effect at higher multipoles.
 In contrast, the effects of a PMF are frequency-independent because
the PMF affects the primary CMB as a background.  Therefore, the
correlation between the PMF and other foreground effects
 should be weak.  Because of this, one should  be able to eventually 
distinguish  the PMF from  foreground effects by using more than two observational data sets at different frequencies.

In summary, we have found that we can constrain more precisely the
power law index $n_\mathrm{B}$ and amplitude $B_\lambda$ of the PMF from all modes of CMB temperature and polarization anisotropies.
The strong
degeneracy of these parameters \cite{Yamazaki:2004vq,Yamazaki:2006bq} is broken by the
different effects of the PMF on the CMB power spectrum for lower and higher $\ell$.
The scalar mode from the PMF can be a main source for lower $\ell$,
while the vector mode can dominate for higher $\ell$ in the CMB
temperature anisotropies.
Furthermore, these calculations suggest that it is possible to place a limit on the correlation parameters $s^\mathrm{(X)}$ for large negative values of the spectral index.  For example ,  
$s^\mathrm{(T)}, s^\mathrm{(S)}< 0$ for $n_\mathrm{B} = -2.9$ is ruled out from the effects of the TT mode on both the lowest and highest multipoles as shown in panels 2a and 2b of Figure 2.   Such results may constrain models for  the origin
of the PMF, along with other PMF parameters. 

\begin{acknowledgments}
We acknowledge Drs. K. Umezu, and H. Hanayama for their valuable discussions.
D.G.Y. and K. I. acknowledge the support by Grants-in-Aid for JSPS Fellows.
This work has been supported in part by Grants-in-Aid for Scientific
Research (17540275) of the Ministry of Education, Culture, Sports,
Science and Technology of Japan, and the Mitsubishi Foundation.  This
work is also supported by the JSPS Core-to-Core Program, International
Research Network for Exotic Femto Systems (EFES).  Work at UND supported in part by the US Department of Energy under research grant DE-FG02-95-ER40934.
\end{acknowledgments}
\begin{figure*}[h]
\includegraphics[width=1.0\textwidth]{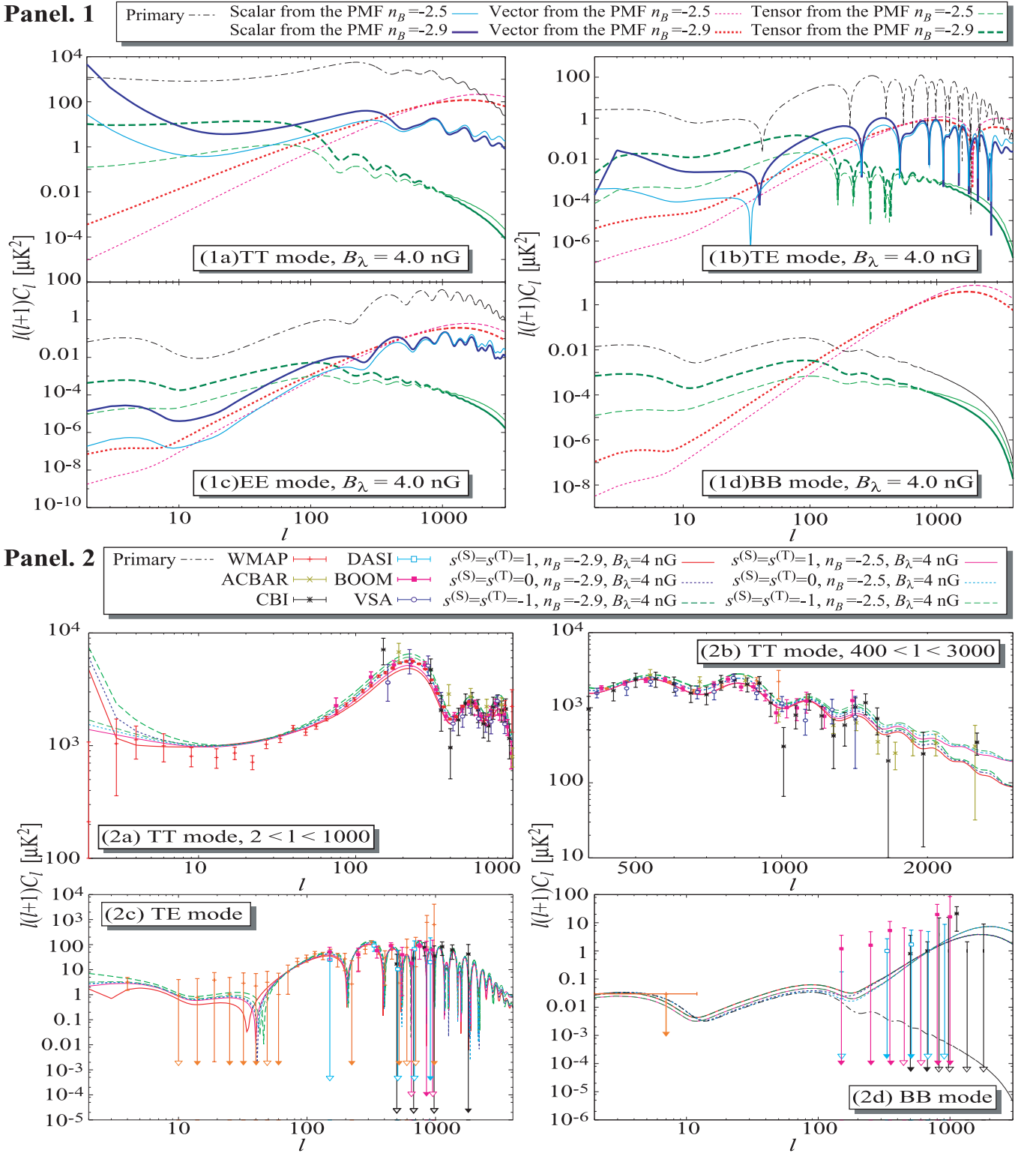}
\caption{\label{fig2}
CMB temperature and polarization anisotropies from the PMF. Panels (1a), (1b), (1c), and (1d)  
show TT, TE, EE, and BB modes, respectively, for models
with $B_\lambda = 4.0$nG and $n_\mathrm{B}$ = $-2.5$ or $-2.9$ as labeled. 
Panels (2a-d) show a comparison of the computed total power spectrum with the observed CMB spectrum for $B_\lambda = 4.0$ nG and various values of $n_\mathrm{B}$, $s^\mathrm{(S)}, and s^\mathrm{(T)}$ as labeled.  Plots show various ranges for: a)  TT($2 < \ell < 1000$), b) TT($400 < \ell < 3000$), c) TE($2 < \ell < 4000$), and d) BB($2 < \ell < 3000$) modes
Curves in all Panels are theoretical lines as indicated in the legend box on the figure.  Lines in (1b) and (2c) are plotted in the absolute value.
Downward arrows for the error bars of Panels (2c) and (2d) indicate that the data points are positive and the lower error negative.
}
\end{figure*}
\bibliographystyle{apsrev2}

\end{document}